# Ro-vibrational dynamics of $N_2O$ in superfluid $^4He$ nano-droplets

Samrat Dey*^, Ashok K. Jha* and Yatendra S. Jain*
*Department of Physics, North-Eastern Hill University,
Shillong-794022, Meghalaya, India.
^Assam Don Bosco University, Guwahati-17, India.

In this paper we use the important interferences of Macro-Orbital theory of superfluidity clubbed with several factors that can change the rotational constant (B) and vibrational frequency of $N_2O$ in $^4He_N$-$N_2O$ clusters with $N$ to account for the results of their recently reported spectroscopic studies which conclude that: (i) in spite of the fact that $^4He$ atoms provide an interactive medium, the rotational spectrum of the clusters shows sharp peaks, similar to that of the molecule in gaseous state, indicating that the molecule rotates like a free rotor, (ii) B decreases monotonically from $N$ = 1 to 6, remains nearly constant for $N$ = 6 to 8 and increases for $N$ = 9 to 10 with small oscillations there after and (iii) vibrational frequency exhibits blue shift for $N$ = 1 to 5 and a red shift for higher $N$.

## 1. Introduction

Superfluid He represents a manifestation of quantum effects at macroscopic level. Numerous studies [1-15] on high resolution ro-vibrational spectra of widely different embedded molecules have been reported in view of the recent finding [3, 4] that the spectroscopy of doped molecules in superfluid He and its nano-droplets (number of He atoms, $N \sim 10^4$) and clusters ($He_N$-M, M being the embedded molecule and $N$ = 1, 2, …) can provide a unique method to explore superfluidity in microscopic systems. It is found that the rotational spectrum of an embedded molecule in He shows sharp peaks, similar to that of the molecule in gaseous state [1-3] which indicates that the molecule rotates like a free rotor, although with an increased moment of inertia ($I$). While, B assumes a new value decreased by a small amount from the gas phase value in case of lighter molecules with B > 1 $cm^{-1}$, it gets reduced typically by a factor of 3-4 for heavier molecules [1-15], indicating that a few He atoms in its surroundings form a part of the rotor. In addition, one also observes a non-trivial dependence of B and vibrational frequency shift on $N$ whose details depend on the constitution of the embedded molecule. We note that different theoretical models attribute all these aspects to the superfluidity of He atoms with different possible assumptions but with little success in achieving a quantitative agreement between theory and experiment. Consequently, in this article, we use a model based on Macro-Orbital theory of $^4He$ [16] and several other factors that affect $I$ and vibrational frequency to explain the experimentally observed ro-vibrational spectra of $He_N$-$N_2O$ clusters, taken as an example.

As mentioned in [2], while, the first spectroscopic experiment in which different atoms were implanted into the interior of liquid He was reported in 1985 and the first similar study of molecule with low resolution was reported in 1992 and with high resolution in 1995 [4], the first infrared spectra of the weakly bound $N_2O$-$^4He$ and $N_2O$-$^3He$ complexes, present in pulsed supersonic jet, were observed using a tunable diode laser in 2002 [10]. The spectra of $N_2O$-$^3He$ were believed [10] to arise mainly from a-type ($\Delta k = 0$) transitions, since, b-type ($\Delta k = \pm 1$) transitions were observed to have very weak intensities. However, both a- and b-type transitions were found to assume good intensities for $N_2O$-$^4He$ complex. The observed rotational parameters were ascertained to agree with an approximately T-shaped structure having: (i) $N_2O$-He bond length ~ 3.4–3.5 Å and (ii) angle between the He position and $N_2O$ axis ~ 80°. The vibrational band origins were found to be slightly blue shifted from those of the free molecule.

Xu et al. [11] reported infrared and microwave spectra of small $N$ (= 1 to 12) $He_N$-$N_2O$ clusters. They found that B values fall with increasing $N$ (up to $N$ = 6) with a subsequent rise from $N$ = 8. Song et al. [12] reported pure rotational transitions of He-$N_2O$ complex and three minor isotopomers (He-$^{14}N^{15}NO$, He-$^{15}N^{14}NO$, and He-$^{15}N^{15}NO$) in the frequency region of 6 to 20 GHz. In case of $^{14}N$ containing isotopomers, nuclear quadrupole hyperfine structure of the rotational transitions was observed and analyzed. The resulting spectroscopic parameters were used to determine geometrical and dynamical information about the complex. A seemingly comprehensive theoretical and experimental analysis of the ro-vibrational spectrum of the He$_N$-N$_2$O complex with $^{14}$N$^{14}$N$^{16}$O and $^{15}$N$^{14}$N$^{16}$O species was reported in [13]. In the year 2006, studies of microwave spectra were extended up to $N = 20$ using $^{15}$N$^{14}$N$^{16}$O species [14], providing a direct experimental evidence of oscillatory change in B with $N$.

Although, these reports [10-14] give the first experimental proof of free rotation of N$_2$O in He clusters, our analysis uses very recent experimental results of Mckellar (published in 2007 [15]) who studied clusters, produced in pulsed supersonic jet expansion from a cooled jet nozzle (> -150 C) with high backing pressure (< 40 atm) and very dilute mixture of N$_2$O (< 0.01%) in He. $^{14}$N$^{14}$N$^{16}$O, $^{15}$N$^{14}$N$^{16}$O and $^{15}$N$^{15}$N$^{16}$O were studied to support the spectral analysis.

While the salient aspects of the Mckellar's paper are given in Section 2, the important aspects of the general theoretical models reported in the literature are summarized in Section 3. We also discuss the prominent aspects of our model and report our findings in Section 4 with conclusion of this study in Section 5.

## 2. Salient Aspects of Experimental Observations

The $N$ dependence of experimentally observed B and the shift of $\nu_1$ frequency as reported in [15] are shown in Fig.1 and Fig.2 respectively.

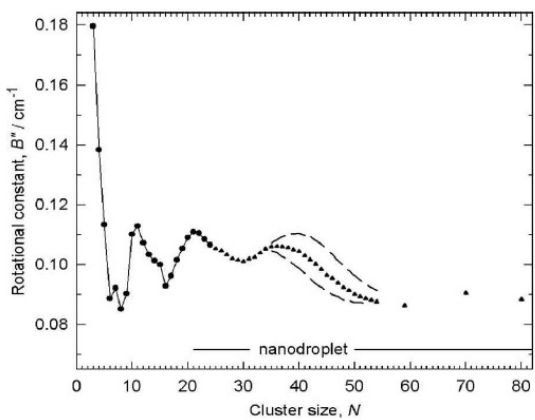 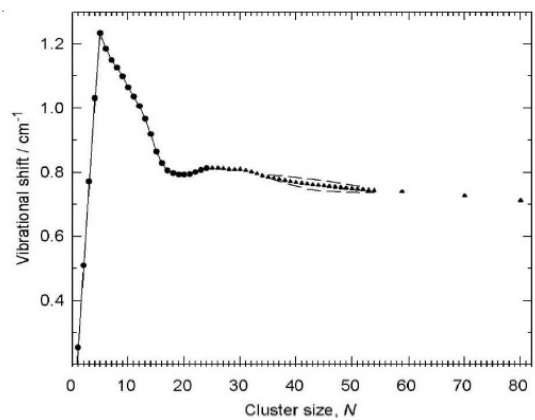

Figure 1: Variation of B with $N$ for He$_N$-N$_2$O clusters (after [15]).

Figure 2: Variation of the vibrational shift with $N$ for He$_N$-N$_2$O clusters (after [15]).

As shown in Fig.1, B decreases monotonically with $N$ increasing from 1 to 6. While it remains nearly constant for $N = 6$ to 8, it increases and decreases in a periodic manner for further increase in $N$. Similarly, vibrational frequency exhibits blue shift for $N = 1$ to 5 and a red shift for higher $N$ with a change in slope at $N \sim 17$. Mckellar also infers that: (i) an unlikely combination of vibrational and rotational effects causes the spectral lines to be lost in unresolved lumps for $N = 23$ to 33 indicating a lesser accuracy in this range, (ii) the fact that B(N$_2$O) < B(CO$_2$), indicates that even for similar molecules, decrease in B for He nano-droplets is not quasi-universal. (iii) the ratio, B"($^{14}$N$^{14}$N$^{16}$O)/ B"($^{15}$N$^{15}$N$^{16}$O), starts out at 1.035 at $N = 0$ (free N$_2$O molecule) and drops to 1.009 at $N = 5$ before climbing smoothly back to values around 1.035 for $N > 10$, (iv) positive values of B'-B" observed in the nono-droplet limit is interpreted to indicate that the molecule rotates slightly more freely in a vibrationally excited state, implying that the N$_2$O-He intermolecular surface is less anisotropic for excited state, (v) small change in molecular centre of mass due to isotropic substitution does not alter significantly the rotational freedom of N$_2$O in He clusters, unlike

He$_N$-CO where such effects have been noted, (iv) small blue shift is observed for *N* = 20 to 25. (vii) A comparison with other molecules shows that while, CO$_2$ and OCS retain 39% and 36% of their free rotational constants, N$_2$O retains only about 33%, (viii) unlike CO$_2$ and OCS, the minimum in B *vs N* remains above the nano-droplet limit for N$_2$O (Fig.1) and the curve in the region of *N* = 4 to 20 is relatively less smooth (although the irregularities of the same magnitude are not observed for vibrational shift); it is suggested that the He-N$_2$O interaction has higher degree of angular anisotropy than that possessed by other two systems. This means that there are more pronounced structural changes in the He$_N$-N$_2$O clusters as a function of *N*.
.
## 3. Theoretical Models

A few interesting models, applicable to any embedded molecule in general, have been used to explain the above mentioned experimental findings. The one known as "supermolecule model" [4], assumes that a specific number of He atoms rotate rigidly with the molecule leading to an increase in *I*. The second model referred to as "two fluid model" [3], takes note of Landau's two fluid phenomenology and relates the observation of free rotation of the embedded molecule with superfluidity of He; it assumes that the density of He atoms around the molecule can be partitioned into spatially dependent normal fluid and superfluid fraction; the former, rotating rigidly with the molecule, has a large value in the first solvation layer. Thus, only the normal fluid fraction contributes to *I*. A slightly different version of two fluid model [1, 17] introduces a new concept of "local superfluid density" and "nonsuperfluid density" (different from normal fluid as envisaged by Landau) and assumes that it is the nonsuperfluid density that rotates rigidly with the embedded molecule. Yet another model, known as "quantum hydrodynamic model" [17-20], adiabatically separates He motion from molecular rotation assuming that the He density in the frame rotating with the molecule is constant and it is equal to that around a static molecule. None of these models, however, successfully explain the experimental observations and add confusions to our understanding of superfluidity. For example: (1) He atoms around the embedded molecules are localized in a very tiny space (of the order of intermolecular distances) where it is impossible to imagine any p=0 condensate [16] and this does not agree with our conventional picture of superfluidity, (2) that the first solvation layer has maximum contribution to normal fluid component which is considered to be responsible for the increase in *I*, does not explain the experimental fact that the light rotors are found to have almost no change in *I* [6], (3) normal fluid density is conventionally associated to the inertial mass of the quasiparticle excitations [1, 21] while such excitations hardly exists in He droplets. Thus, as stated in [1, 2, 8], we indeed lack a viable theory that comprehensively explains different aspects of experimental observations.

## 4. Analysis Based on Macro-Orbital Theory

According to Macro-Orbital (MO) theory of $^4$He, superfluidity arises due to a localized and orderly configuration of He atoms viz. identically equal separation (=$\lambda$/2, $\lambda$ is the de-brogle wavelength) in normal space with a phase difference of 2n$\pi$ in phase space and identically equal value of zero-point momentum ($q_o$ = $\pi$/d, d being the inter-particle distance) in momentum space. It further concludes that while particles are localized and loose their relative motion, they are free to move in the order of their location on a closed path. The model reveals that there is no question of partial depletion of condensate for strongly interacting bosons (e.g. He atoms); instead, almost all atoms below $\lambda$-point assume their ground state where they have identically equal energy, indicating that all atoms leave q = 0 state to occupy q = q$_o$ state. Consequently, we use a localized and orderly arrangement of He atoms to explain the experimental findings.

It is well known that each He atom added to a cluster, He$_N$-M, can render new bond lengths and bond angles providing a different structure and symmetry of He$_{N+1}$-M which obviously has different *I*. However, it is seen that this simple fact is not sufficient to account for experimentally observed decrease in *I* with increasing *N*. As such, we consider that those He atoms in the cluster which happen to be in the molecular plane render a sort of an equi-potential ring for a change in the angular posture of the rest of the cluster which can therefore rotate like a free rotor. This consideration finds strong support from the fact that He atoms in different rings in a quantum vortex move with different speeds without interfering with each other [21]. Thus, each He atom joining other atoms in the said equi-potential ring does not increase *I*. On the contrary, as the number of such He atoms increases, the order of the rotational symmetry (n) of the potential, $V_n = (V_0/2)[1-\cos(n\emptyset)]$, increases which tends to decrease *I*, as evident from the analysis Matthew's equation [22]; we assume $V_0 \ll 2B$, in conformity with the observed free rotation. In fact, Mathew's equation also suggests that the increase in the height of the potential hills, $V_0$ ($\ll 2B$), also decreases *I*. We use these inferences to explain the experimental observations.

Identifying the molecular axis with y axis and that the molecule rotates about x axis, He atoms are first considered to occupy positions on a ring (called as the 'first ring') in x-z plane. These atoms have reasonably good binding with the molecule and the resulting complex behaves like a rigid rotor, as shown in Fig.3 and Fig.4, for $N$ = 3 and 6 respectively. Obviously, *I* of the rotor increases with increase in *N* of such atoms. From the fact that B assumes minimum value for $N$ = 6 (Fig.1) and the blue shift in $v_1$ assumes maximum value around the same *N* (Fig.2), it appears that the first ring gets saturated with about 6 He atoms. Further, the fact that for $N$ = 7 and 8 there is no significant change in B, indicates that those extra He atoms occupy positions (say, X, which can be identified with the centre(s) of the gray sphere(s) attached to $N_2O$ by longer bond(s) in Fig.5) close to the principal moment of inertia axis about which the complex rotates. Although, B for $N = 9$ has almost the same value as that for $N = 7$ and 8, a third He atom on the axis of rotation would be a second nearest neighbor, while, there are several positions available for the He atom to occupy as the first nearest atom to the molecule. It, therefore, appears that the addition of ninth He atom renders a complex with a new structure (presumed to be identified as [5+4] structure) in which only 5 He atoms remain in the first ring, while 4 He atoms constitute a 'second ring' (the said equi-potential ring) in x-y plane. Obviously, this ring does not rotate with the rotor ($He_5-N_2O$) and hardly makes significant contribution to $I$; a similar arrangement envisaged for $N = 10$ is shown in Fig.6 for better perception of the rotor and the second ring. As, there is a sharp increase in B with $N$ changing from 9 to 10, it is reasonable to assume that there is a decrease of one He atom in the rotor part of $N = 10$ complex which means that the complex has [4+6] structure of which only 4 He atoms contribute to $I$. Since, B changes only slowly with $N > 10$ (Fig.1), it appears that the extra He atoms go to the second ring rather than becoming a part of the rotor. However, the extra He atoms keep on changing the structure of the second ring which can obviously influence the first ring by modifying the He-$N_2O$ bond length and/or its planar structure in such a way that $I$ of the rotor increases smoothly with $N$ (12 to 16) (Fig.1) beyond which it has a smooth decrease followed by a kind of its oscillatory change with $N$. Similarly one may also observe that there is a change in slope for red shift in $v_1$ around $N = 16$ (Fig.2). Both these observations imply that the second ring gets saturated with about 12 (= 16 - 4) He atoms. It may be noted that the second ring can accommodate more He atoms than the first ring because it surrounds the length of the molecule and not the molecular axis. When He atoms have occupied all the possible sites (apparently, about 16) close to these two rings, one may visualize the formation of a 3-D saturated shell by these atoms. Any further addition of He atoms makes a beginning of second shell which should also get saturated for certain number of He atoms leading to the beginning of third shell and so on. Since the evolution of these shells can make a periodic change in the structure of the rotor with change in $N$ we rightly observe the oscillation in B (Fig.1) and corresponding indicators in the shift of $v_1$ frequency (Fig.2). One can find that these arguments help in having a good account of the experimental observations (Fig.1 and Fig.2). This will be all the more evident from the following.

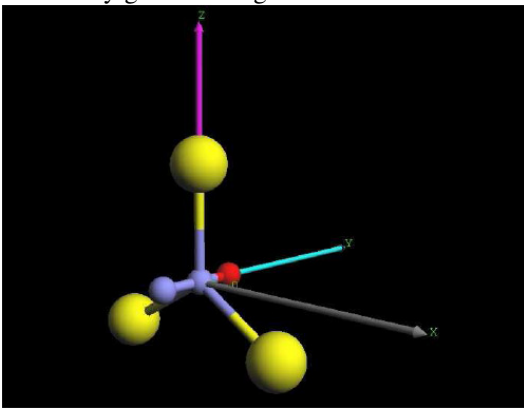

Figure 3: Three He atoms are attached to the molecule and rotating with it.

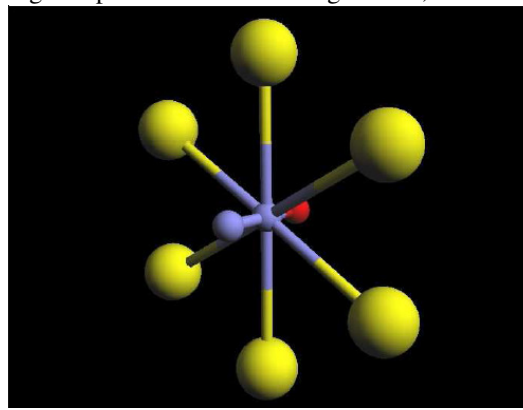

Figure 4: Six He atoms are attached to the molecule and rotating with it.

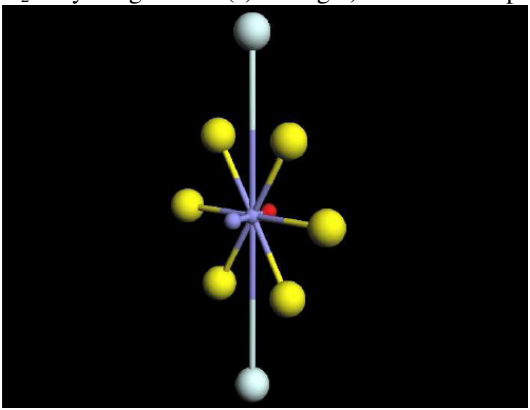

Figure 5: Six He atoms are rotating with the molecule and two He atoms fall in the moment of inertia axis.

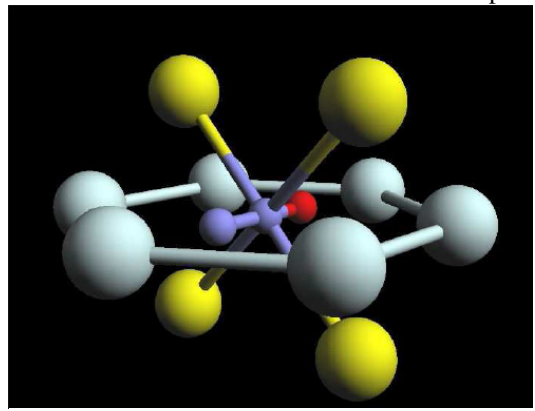

Figure 6: Four He atoms are attached to the molecule and six He atoms providing the equi-potential ring.

Fig.7 and Fig.8 show the required He-$N_2O$ distance 'D' in the rotor (calculated by using a computer program [23]), which reproduces the experimental value of B for different $N$. From Fig.7 we find that D increases for $N = 3$ to 6 which is consistent with the fact that strength of He-$N_2O$ interaction decreases with increasing number of He atoms attached to the molecule. Again from $N \sim 17$ onwards (Fig.8) we observe fluctuations in D which is also quite justified as we consider it to be because of the addition of successive shells. For $N = 7-8$, we do not expect D to differ from that obtained for $N = 6$ (since 7[th] and 8[th] He atoms do not contribute to $I$, as discussed above), while, for $N = 9$ where the rotor has 5 He atoms, either D is expected to increase or the structure of the rotor is expected to have more angular deformation, due to the presence of 4 atoms in the second ring. For $N = 10$ to 16, we consider two options: (i) change in D with no angular deformation in the rotor and (ii) change in the angular positions '$\theta$' of He atoms in the rotor without any change in D; it may be noted that the rotor has only 4 He atoms. Here '$\theta$' means the angle by which the He atoms of the first ring are shifted away (along ±y direction) from its planer structure (z-x plane), which reproduces the experimental value of B for different $N$ (calculated by a computer program [24], similar to [23]). Possible values of D and $\theta$ obtained by using the said options are depicted as a function of $N$ in Fig.9 and 10, respectively. While, the range of possible D and $\theta$ are within the expected range, the systematic increase in D or $\theta$ of the rotor with $N$ (10-16) seems to arise due to the interaction of the He atoms of the second ring with those of the rotor. Noting that the length of $N_2O$ molecule (2.312 Å) is small (compared to the lengths of molecules like OCS (2.718 Å) etc.), implying that the second ring of He atoms around the rotor is also small, He atoms of the second ring are expected to reach fairly close to the He atoms of the rotor. Consequently, He-$N_2O$ bond length and/or angle in the rotor can change significantly with $N$ (~ 10-16) and thereby increase $I$ appreciably. Apparently, this +ve change in $I$ is much larger than its –ve change resulting from $V_0$ and the order of symmetry (n) of the potential, $V_n = (V_0/2)[1-\cos(n\emptyset)]$.

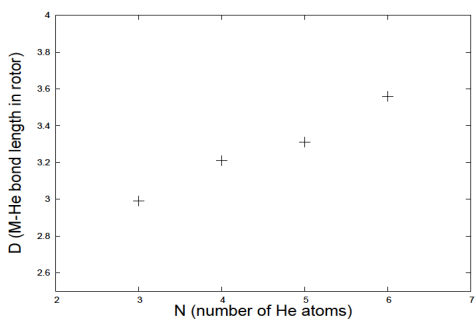
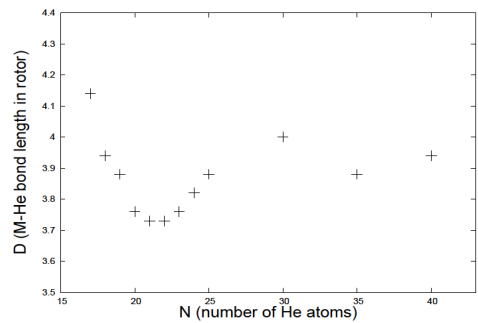

Figure 7: Change of D (in Å) with $N$ for $N = 3 - 6$

Figure 8: Change of D (in Å) with $N$ for $N = 21 - 40$

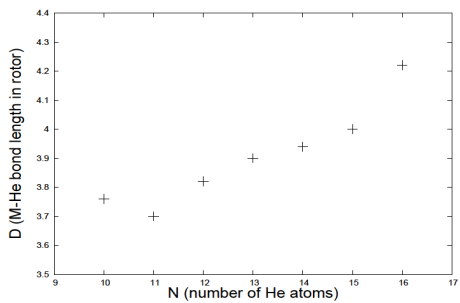
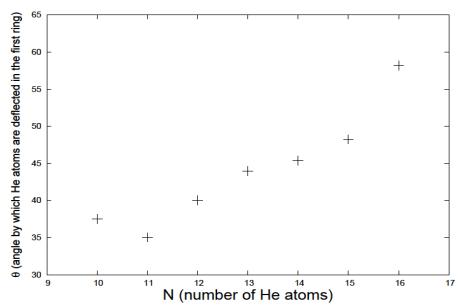

Figure 9: Change of D (in Å) with $N$ for $N = 10 - 16$

Figure 10: Change of $\theta$ (in degrees) with $N$ for $N = 10 - 16$

## 5. Conclusions

We note that: (i) The small magnitude of total blue shift (which gets even smaller with increase in $N$) indicates that He-$N_2O$ interaction is very weak. (ii) The nature of the spectroscopic ro-vibrational structure of the embedded molecule is similar to that of the molecule in gas phase [1, 2] which indicates that He atoms around the molecule to a good approximation render isotropic potential for rotation of the rotor. (iii) The sharp rotational and vibrational spectral lines [1-3] indicate the fact that He atoms in He$_N$-$N_2O$ clusters do not have relative motion with respect to the rotor or vibrating $N_2O$; in other words, all He atoms get localized (with zero-point uncertainty in position) and form a structure which does not change with time, as only this ensures that the rotor or the vibrating molecule sees a time independent potential, necessary for sharp spectral lines; to a good approximation, this requirement is satisfied even if the particles around the rotor move in the order of their locations. These observations are consistent with all the relevant conclusions of MO theory (Section 4) and this justifies its use in accounting for the experimental observations for the change of B and vibrational shift with $N$. As such, we basically consider different orderly and localized arrangements of He atoms, having different He-$N_2O$ bond lengths and bond angles, for the said changes. It is also noted that He atom occupies a space as per its energy of localization; no two helium atoms come closer than their wave packet size. We find that our approach based on the MO theory explains the experimental results to a good accuracy. It may be mentioned that our present calculations do not consider, quantitatively, the change in $I$ due to the increase in order of symmetry (n) of $V_n$ and/or the change in the height of the potential hills ($V_0$). We plan to consider these factors in our future course of calculations which will be reported at a later date.